# Elementary Intracellular Ca Signals are Initiated by a Transition of Release Channel System from a Metastable State

**Guillermo Veron[1], Victor A. Maltsev[1], Michael D. Stern[1], Anna V. Maltsev[2]***

[1]Cellular Biophysics Section, Laboratory of Cardiovascular Science,

National Institute on Aging, NIH, Baltimore, MD 21224, USA

[2]School of Mathematical Sciences, Queen Mary University of London, London E14NS, United Kingdom

***Corresponding author:**

Anna V. Maltsev, PhD
Email: a.maltsev@qmul.ac.uk
Telephone: +44 (0)20 7882 2969

## Abstract

Cardiac muscle contraction is initiated by an elementary Ca signal (called Ca spark) which is achieved by collective action of Ca release channels in a cluster. The mechanism of this synchronization remains uncertain. This paper approaches Ca spark activation as an emergent phenomenon of an interactive system of release channels. We construct a Markov chain that applies an Ising model formalism to such release channel clusters and realistic open channel configurations to demonstrate that spark activation is described as a system transition from a metastable to an absorbing state, analogous to the pressure required to overcome surface tension in bubble formation. This yields quantitative estimates of the spark generation probability as a function of various system parameters. Our model of the release channel system yields similar results for the sarcoplasmic reticulum Ca concentration threshold for spark activation as previous experimental results, providing a mechanistic explanation of the spark initiation. Additionally, we perform numerical simulations to find spark probabilities as a function of sarcoplasmic reticulum Ca concentration obtaining similar values for spark activation threshold as our analytic model, as well as those reported in experimental studies.

## Introduction

Robust intracellular signals are achieved by synchronous operation of groups of molecules, each operating stochastically. In cardiac muscle, Ca release channels, ryanodine receptors (RyR), form clusters of 20 to 200 channels (known as Ca release units, CRU) embedded in the membrane of the junctional sarcoplasmic reticulum (SR). The channels can be opened via a stochastic gating mechanism that is activated by Ca and then close spontaneously. Within CRUs, individual channels are very close to each other (~30 nm) and arranged in an almost perfect rectangular grid. The channels interact via Ca-induced-Ca-release (Fabiato, 1983) thereby facilitating RyR openings throughout the cluster. The all-or-none event when almost all of the channels in a CRU have been opened is referred to as a Ca spark (Cheng et al., 1993;Cheng and Lederer, 2008).

Ca sparks can be triggered by a Ca influx via L-type Ca channel ("induced sparks"), or arise spontaneously ("spontaneous sparks" (Cheng et al., 1993)). Induced sparks are elementary signals of excitation-contraction coupling and spontaneous sparks contribute to normal cardiac impulse initiation in sinoatrial node cells (Lakatta et al., 2010). Thus, spark activation is of crucial physiological importance in cardiac muscle contraction as well as heart rate and rhythm. Under pathological conditions spontaneous sparks can create an excessive amount of Ca leaking from the SR that has deleterious effects on heart function. In ventricular myocytes spontaneous sparks can initiate Ca waves and life-threatening arrhythmia (Bers, 2014;Boyden and Smith, 2018). In all these circumstances understanding spark initiation is essential both for basic biophysical research as well as in biomedical applications.

Here we study under what conditions RyRs open simultaneously to create a full Ca spark instead of firing individually or partially synchronized, all of which have been observed experimentally under various conditions (Wang et al., 2004;Brochet et al., 2011). Zima et al. (Zima et al., 2010) found that full sparks start forming as the SR Ca load surpasses 300 μM. Despite its fundamental importance, spark activation has not been systematically studied theoretically or numerically as an emergent phenomenon of an interactive system of release channels.

Detailed numerical models of the entire CRU including individual, interacting, stochastically gated RyR channels in the dyadic space were reported by Laver et al. (Laver et al., 2013) and Stern et al. (Stern et al., 2013), focusing on Ca spark termination. This approach was extended towards understanding the effect of different CRU geometries (Walker et al., 2014). Other models approximate CRU phenomenologically by a single gating mechanism or as a Markov chain representing a result of interactions of all RyRs within the CRU (the "sticky cluster" model (Sobie et al., 2002)). In 2011, Sato and Bers approximated probabilities of different number of RyRs open in the CRU at a given junctional SR Ca level by using the binomial distribution (Sato and Bers, 2011). However, due to the assumption of independence for RyRs inherent in the binomial distribution, this approach lacks the effect of RyR interactions crucial for spark initiation.

A fundamentally new approach to describe CRU operation has recently been introduced by the authors in (Maltsev et al., 2017), where the Stern model of the CRU (Stern et al., 2013) was mapped isomorphically to the Ising model from statistical mechanics. Further analysis identified the critical parameter (referred to as β, similar to inverse temperature in statistical mechanics models) that determines conditions for Ca leak (Maltsev et al., 2019). Both these studies focused again mainly on spark termination. The present study is the first application of the Ising formalism to spark activation. Here we introduce a new Markov chain describing the numbers of adjacent open channels to explicitly estimate the probability that an open RyR will develop into a spark at each level of SR Ca, thus establishing the threshold SR Ca load at which a spark can occur and offering a mechanistic explanation. For numerical approximation, here we employed the modified Stern model of Ca spark. In addition to our analytic approach, the present work is the first numerical study to shed light on activation in a cluster of interactive RyRs and to establish a threshold of SR Ca loading for spark activation.

## Methods

***Background.*** The Stern numerical model of the CRU has been shown to be isomorphic to an Ising model (Maltsev et al., 2017), a classical model of statistical physics used to explain spontaneous magnetization. This isomorphism provides a starting point for the present work. The RyRs are assumed to be in a rectangular grid (with their coordinates given as $x = (x_1, x_2)$ and with (0, 0) as the center of a grid with an odd number of elements) with each RyR assuming one of two states: open (+1) or closed (-1). An assignment σ of an open or closed state to each RyR is called a configuration, and the Ising model is a continuous time Markov chain with RyR configurations as states. The instantaneous transition rates are only non-zero between configurations differing at only one RyR, say at position *x*, and, upon discretizing time, are given by Eq.6 in (Maltsev et al., 2017), which we give here for convenience:

$$P(\sigma,\sigma^x) = \begin{cases} \Delta t C e^{2\beta(\sum_{y\in\Lambda_b}\phi(|x-y|)\sigma(y)+h)} & \text{for } \sigma(x)=-1 \\ \Delta t C & \text{for } \sigma(x)=1 \end{cases} \quad (1)$$

Here $\sigma^x$ is the configuration that coincides with configuration σ except at *x* where the state is reversed and $\phi(r)$ is the interaction profile (defined below).

The closing rate C is taken to be constant, and the opening rate is taken to be an exponential of the cleft Ca concentration given by λ* exp(γ[Ca]) fitted to experimental data of Laver et al. (Laver et al., 2013) in our previous study (Maltsev et al., 2017). The analogues of magnetic field *h* and inverse temperature β are given by $\beta = \frac{\gamma\psi(U)}{4}$ and

$$h = \frac{1}{2\beta}\ln\frac{\lambda}{C} + \sum_{\substack{x\neq 0 \\ x\in CRU}} \phi(|x|) \quad (2)$$

where *U* is the distance between RyRs, $\psi$ is the Ca level in the cleft resulting from the opening of an RyR as a function of distance from the open RyR (i.e. an interaction profile, Figure 1), and $\phi(r) = \psi(rU)/(\psi(U))$ is a natural choice of scaling for the interaction profile function $\phi$.

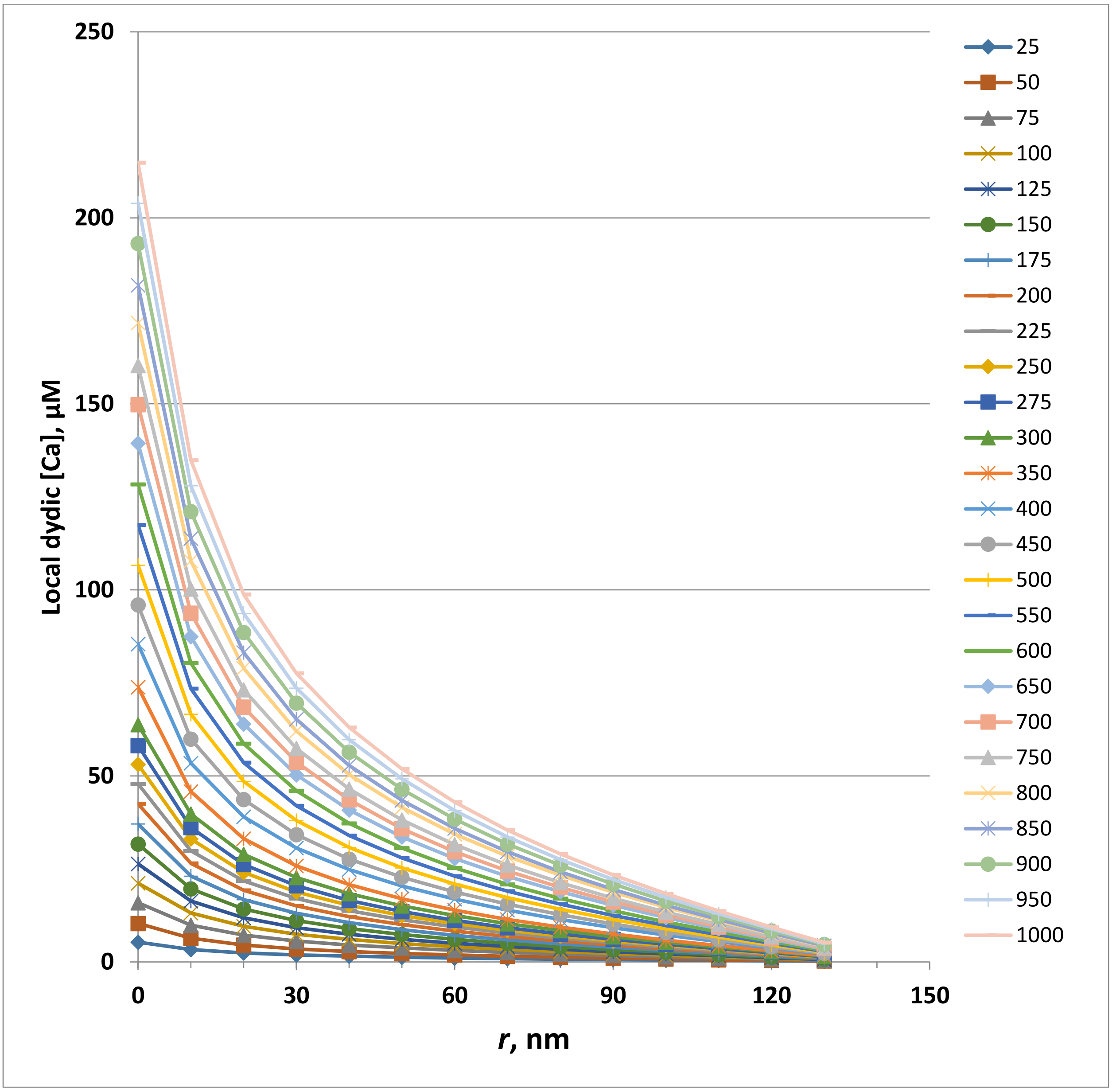

**Figure 1**: A family of interaction profiles *ψ(r)* at various SR Ca. *ψ(r)* is a steady-state [Ca] in the dyadic cleft as a function of distance *r* when one RyR opens at *r*=0. The profiles were measured in numerical simulations for 9x9 RyR grid, the distance between RyRs is 30 nm, and each voxel is 10x10x15 nm in xyz. See supplemental Excel file for exact values of the profiles.

***The new Markov chain.*** We follow an evolution of a cluster under the conditions of strong interactions (i.e. supercritical *β*) and favorable magnetic field (i.e. positive and growing) but an initial configuration where all RyRs are closed (maximally unfavorable). For a wide range of positive magnetic field *h*, this initial condition constitutes a local energy minimum (also known as a metastable state) and the system is highly unlikely to transition to an all-open state. It would require the unlikely event of several RyRs randomly opening next to each other, despite the closed neighbors. Only when *h* is large enough that one open RyR creates enough Ca flux to strongly influence its neighbors, a spark has a good chance of activating.

To quantify these concepts, we introduce a new Markov chain. We define an *open cluster* as a collection of channels that are open and adjacent (diagonals don't count). The states of the Markov chain are the size of the open cluster going from 0 to 4 (Figure 2A) and the transition probability for increasing the cluster is computed from (1), but weighted by the relative frequency of configurations both in the initial and the target states. The transition probability of decreasing the cluster is computed from (1) as well, but we assume that when transitioning from 3 to 2 only the outside RyRs can close, resulting in the configuration as in state 2.

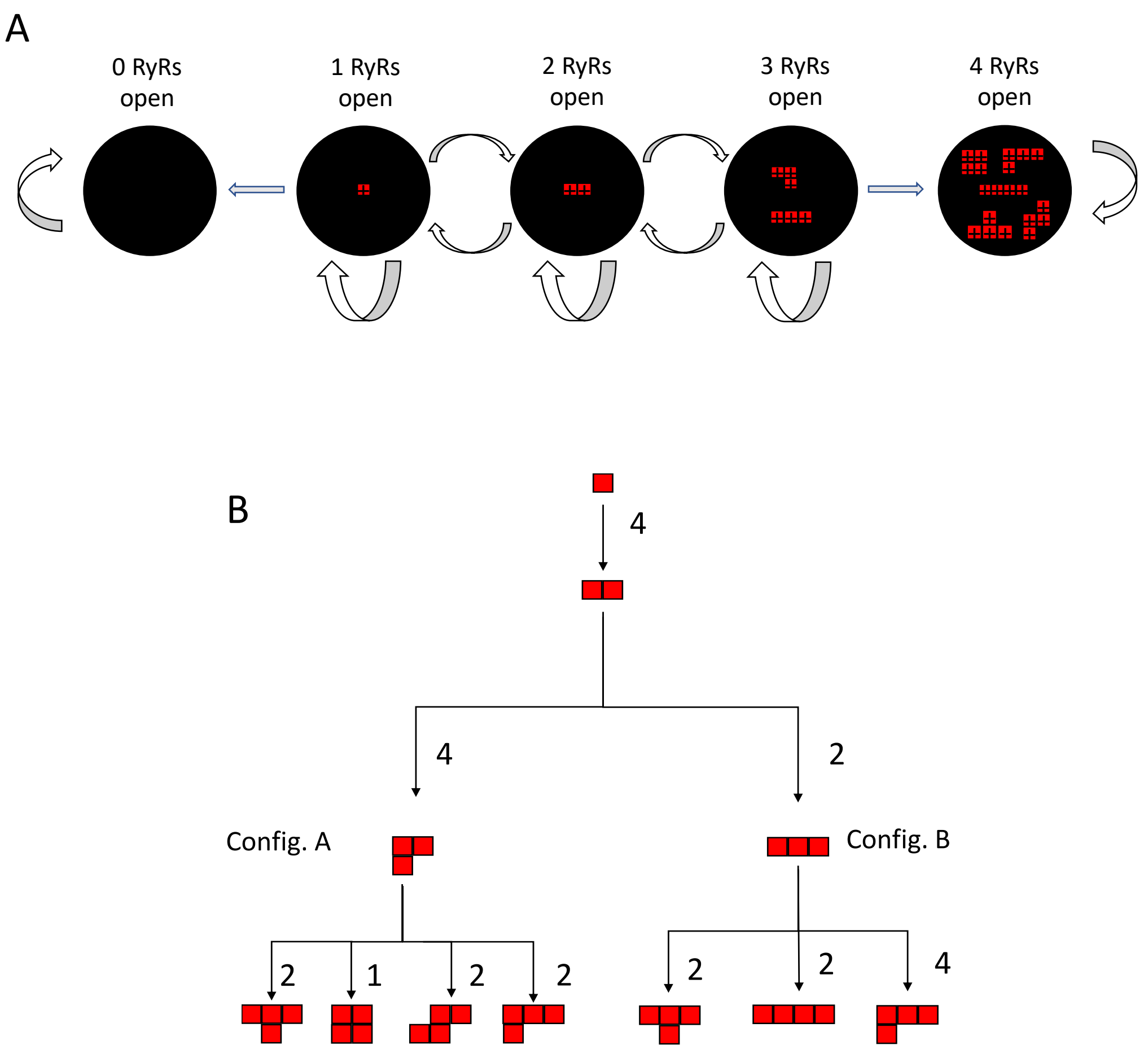

**Figure 2.** Evolution of RyR cluster to generate a spark. **(A),** Five-state Markov process simulating the evolution. Each arrow represents the event of the Markov process changing from one state to another state with the direction indicated by the arrow. (**B),** Configuration tree. Illustration of the different configurations possible and the series of events that could take place to reach each of the configurations. Numbers at each configuration indicate the number of possible ways to reach a given configuration.

## Results

***Calculation of transition probabilities.*** Here we show an example of a calculation for the transition matrix of our Markov process. We compute the probability of going from 1 open channel to 2 open channels $P(1 \rightarrow 2)$ at an SR level of 300 μM. We find the following:

$\beta$ at 300 μM: 0.64544597505
$\Sigma_{y\epsilon\Delta_b}\phi(|x-y|)\sigma(y)$ at 300 μM: -20.789863109249815
h at 300 μM: 18.02130417561286
C (closing channel rate) = 117 $s^{-1}$
$\Delta t = 7 * 10^{-10}$ ms

$$P(1 \rightarrow 2) = \Delta t C e^{2\beta(\Sigma_{y\epsilon\Delta_b}\phi(|x-y|)\sigma(y)+h)}$$
$$= (7 * 10^{-10})(117)e^{2(0.64544597505)(-20.789863109249815+18.02130417561286)}$$
$$P(1 \rightarrow 2) = \Delta t C e^{2\beta(\Sigma_{y\epsilon\Delta_b}\phi(|x-y|)\sigma(y)+h)} = 2.2969646869805702 * 10^{-9}$$

Lastly, since there are four different ways for one open channel to turn into two open (adjacent) channels, we multiply this probability by four to arrive at the final answer of

$$P(1 \rightarrow 2) = 9.187858747922281 * 10^{-9}$$

The calculation becomes more involved for P(2 → 3). We have a formula for transition probability from a given configuration to a configuration with one square added. Looking at Figure 2B, to compute P(2 → 3), we compute the probability of transitioning from a configuration with 2 squares to configuration A (in the left branch) and multiply it by 4 and then add the probability of transitioning from a 2 to configuration B (in the right branch) and multiply by 2. Lastly, to obtain P(3 → 4) we notice that if we pick a random configuration with 3 squares, with probability 2/3 it will be in configuration A and with probability 1/3 it will be in configuration B. Thus to compute P(3 → 4) we find the appropriately weighted sum of probabilities of transitioning from configuration A to a configuration of 4 squares and multiply the result by 2/3 and similarly find the appropriately weighted sum of probabilities of transitioning from configuration B to a configuration of 4 squares and multiply the result by 1/3.

This procedure results in a transition matrix $M = (P(i \rightarrow j))_{0 \le i, j \le 4}$. The canonical form of an absorbing chain, with *t* transient states and *r* absorbing states, is defined as the following:

$$P = \begin{pmatrix} Q & R \\ \mathbf{0} & I_r \end{pmatrix}$$

Where *Q* is a *t*-by-*t* matrix, *R* is a nonzero *t*-by-*r* matrix, *0* is the *r*-by-*t* zero matrix, and $I_r$ is the *r*-by-*r* identity matrix. To obtain a matrix in canonical form, we relabel the state with 0 squares as state 5. The fundamental matrix is defined as the following:

$$N = \sum_{k=0}^{\infty} Q^k = (I_t - Q)^{-1}$$

The probability of transitioning from state *i* to state *j* in exactly *k* steps is the (*i*,*j*) entry in the matrix $Q^k$. Lastly, the absorbing probability for going from a particular transient state *i* to a particular absorbing state *j* is the (*i*, *j*) entry in the matrix *B* = *NR*. Using this formula, we obtain the probability of going from state 1 with one open RyR to the absorbing state 4 with 4 open RyRs by finding the (1, 4) entry in the matrix *NR*. We use Theorems 11.5 and 11.6 of (Grinstead and Snell, 1997) for absorbing Markov chains to obtain the probability of absorption in state 4 when starting from state 1, corresponding to the probability that a spark initiates from a given open RyR (for Python code see Supplemental material).

Figure 3 shows the results of similar computations performed by our Python program for all relevant values of SR Ca yielding the probability of transitioning from 2 open channels to 3 open channels and the probability of transitioning from 3 open channels to 4 open channels (our Python code and its Psi data file are provided in the supplement). We notice that the curve P(3 → 4) is steeper and lies to the left of P(2 → 3), so SR Ca at which the transition from 2 open channels to 3 open channels becomes somewhat likely is the same as SR Ca where the transition from 3 open channels to 4 open channels becomes extremely likely. This indicates the dependence of growth of the open cluster on its size. On a physics level, this happens because the system with all channels closed but positive magnetic field and supercritical $\beta$ is in a local energy minimum. Each individual channel or small cluster might not open or, if open, close quickly due to the strong interaction from closed neighbors. But as the open cluster grows, the "curvature" of its boundary decreases, so the force from the closed neighbors gets distributed over more open neighbors and is less likely to close an open channel.

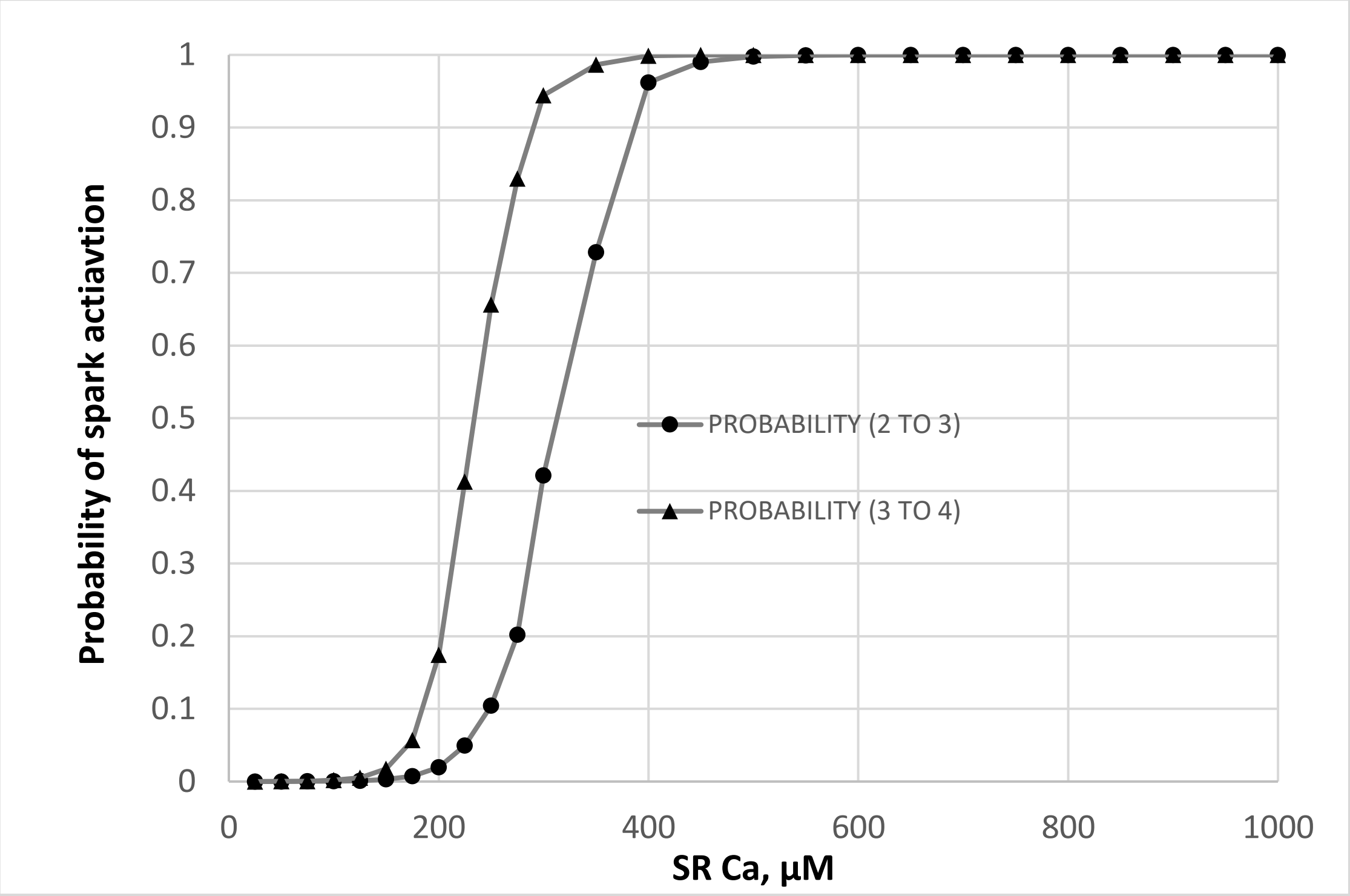

**Figure 3**: The probability of transitioning from 2 open channels to 3 open channels (circles) and probability of transitioning from 3 open channels to 4 open channels (triangles). We notice that the SR Ca at which the transition from 2 open channels to 3 open channels becomes somewhat likely is the same as SR Ca where the transition from 3 open channels to 4 open channels becomes extremely likely.

***Spark activation probability as a function of SR Ca.*** In this work we assume that once the open cluster reaches 4 in size, it always initiates a spark. We do not compute in this work the probability of transitioning from 4 to 5, we call it P(4 → 5), because the enumeration of all the possible clusters becomes cumbersome. However, it is clear that the curve P(4 → 5) as well as further curves such as P(5 → 6) would be much steeper and lie to the left of P(3 → 4), in a similar way as P(3 → 4) is steeper and lies to the left of P(2 → 3) as evidenced in Figure 3. Thus, it is reasonable to let 0 and 4 be absorbent states.

The results of our analytical modeling are shown in Figure 4B (circles). The probability of spark activation follows a steep sigmoid curve as a function of SR Ca load, beginning to rise at around 250 μM. Our analytic results closely match the results of experimental studies (Figure 4A) and our numerical simulations (Figure 4B, triangles). More sensitive spark generation at high SR Ca vs. numerical estimates reflects analytical model assumption of instantaneous interactions, whereas Ca diffusion causes a small delay. In numerical model it takes roughly 2.5 ms for the Ca profile to reach its stable level (Figure 2B of (Maltsev et al., 2017)). Approximating the interactions with a step-function which is 0 until 1.25 ms and the full profile after 1.25 ms, and using the closing rate from our numerical model of C=0.117 $ms^{-1}$, we obtain that with probability of approximately 15% the RyR will close before it has a chance to interact with other channels. On the other hand, with probability 85% it will interact and enter into our Markov chain setup. Thus, if we scale the curve by 0.85 to account for this discrepancy, we obtained a closer match at higher SR Ca (Figure 4C). Less sensitive spark generation at low SR Ca in analytical model can be due to other Ising model assumptions, such as its interactions limited to the nearest neighbors. On the other hand, the threshold of spark activation (300-400 μM) reported in experimental studies (Figure 4A) is better reproduced by our analytical model than by the numerical modeling (200-300 μM).

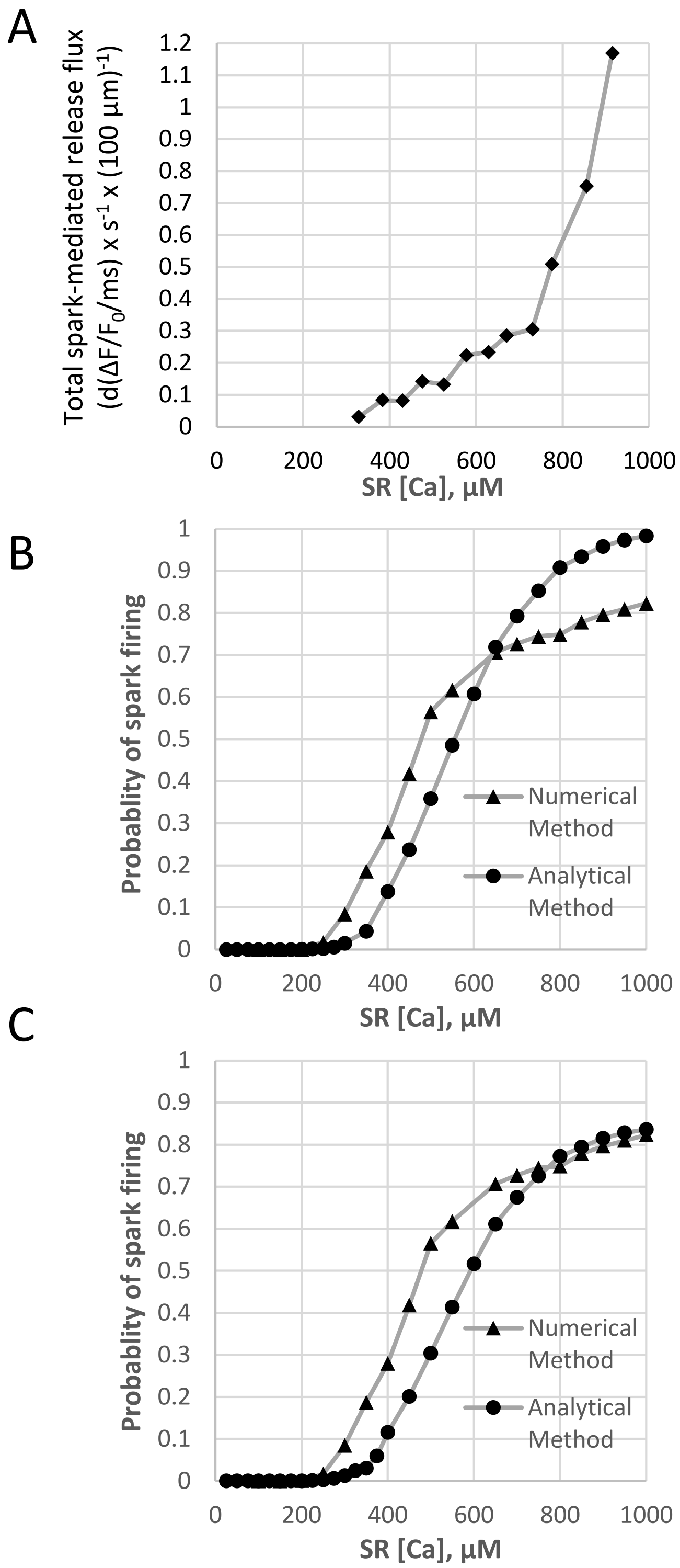

**Figure 4.** Spark activation as a function of SR Ca load. **(A),** Experimentally defined SR Ca threshold for Ca spark activation; shown are mean values of total spark-mediated release flux (measured by confocal microscopy) which were rescanned and replotted from Figure 3B of (Zima et al., 2010). Spark activation predicted numerically and analytically: **(B),** no correction for diffusion delay and **(C),** with correction for the delay. In numerical method, probability of spark firing at each SR Ca was evaluated from 10,000 simulation runs of 200 ms each. In each run at *t*=0 one RyR in the center of 9x9 RyR cluster was set open. Our criterion for spark firing was that 50% of all RyRs open at any moment before all RyRs closed.

## Discussion

In this study, we use numerical and analytical approaches to study Ca spark activation. We consider spark activation as an emerging property of an ensemble of interacting individual RyRs. Using this approach, we predict the threshold of SR Ca required for spark activation based on intrinsic RyR channel properties and RyR interactions. We examined spark activation dynamics at different SR Ca levels and found a sharp transition at 300 μM level where sparks were robustly generated. Zima et al. (Zima et al., 2010) reported spark and non-spark Ca SR leak types in ventricular myocytes. As SR Ca load grows above ~300 μM, Ca sparks strongly contribute to the leak (Figure 4A). Our results are in agreement with previous experimental studies.

The present study provides a mechanistic view on the spark initiation threshold. Mathematically speaking, one can view spark activation as equivalent to a system transitioning from a local energy minimum (also known as a metastable state) to a global one. As a physical process, spark activation becomes analogous to the pressure required to overcome surface tension in bubble formation. When a open cluster forms in a background of closed channels, the interaction between the closed and the open channels happens only at the boundary of the open cluster. When the open cluster is small, there are more closed channels per open channel at the boundary (in the extreme case of one open channel, there are 4 closed channels per that one open channel and the Ca from just one channel gets spread out to all 4). As the open cluster gets bigger, this ratio gets more favorable for the open channels. This reflects the "curvature" of the boundary as in bubble formation.

Our model includes measurable parameters of the system that can be further varied to understand the impact of realistic factors for spark activation, and our theoretical framework allows for further predictions. Such factors include SERCA pumping rate to increase network SR Ca, gating of RyRs and their sensitivity to Ca, connectivity of free SR and junctional SR (Sato et al., 2016), Ca buffering (e.g. via calsequestrin), posttranslational modifications (such as phosphorylation) of key Ca cycling molecules, etc. For example, increasing SR connectivity or performance of SERCA function would bring the system to spark activation threshold quicker as a result of SR Ca reloading at a more rapid pace. If diastole is long enough, these would provide conditions for spontaneous Ca release still during diastole, triggering life-threating arrythmia (Bers, 2008;Dobrev and Wehrens, 2017). On the other hand, the increased diastolic Ca release contributes to normal generation of spontaneous pacemaker potentials driving the heartbeat (Lakatta et al., 2010). Thus, our approach could help in directing pharmacological interventions to avoid regimes of spontaneous spark activation in cardiac muscle cells in cardiac disease (Bers, 2008) or to promote such regimes in cardiac pacemaker cells in sick sinus syndrome (insufficient pacemaker function) (Dobrzynski et al., 2007).

Lastly, our new analytical approach provides a substantial computational advantage to evaluate the conditions for spark activation within Ca release channel clusters. Calculating the dynamics for all states in the full Markovian representation of a CRU using the analytic solution to Markov matrix would involve taking exponentials of enormously large matrices.

## Acknowledgments

The work was supported by the Intramural Research Program of the National Institute on Aging, National Institutes of Health. A.V.M. acknowledges the support of the Royal Society University Research Fellowship UF160569.

Supplementary Python code

This program was created by Guillermo Veron as a part of his research work in 2019-2021 at the Laboratory of Cardiovascular Science, Cellular Biophysics Section, Intramural Research Program, National Institute on Aging, NIH, Biomedical Research Center 251 Bayview Blvd. Baltimore, MD 21224-6825.

# Instructions/Explanations

**'Psi Values.xlsx' is needed in the same folder as code to work properly. Simply run the code and final probabilities of spark occurring will be listed in "A_LIST." This will be found under the "Diagonalization" sub-section under "Probabilities" section**

***Global Variables***

**We establish our data and constants we will be working with. For example, the number of nodes, psi values, given SR levels, value of gamma, channel closing rate, etc.**

***Hamiltonian Functions***

**We create our hamiltonian variables and functions such as how to calculate phi values, interaction profile values, and magnetic field values.**
**The interaction profile values will necessarily change for different configurations. For example, going from one open node to two open nodes, going from three open nodes (in a straight configuration) to four open nodes (in an "L" configuration), etc. All potential configuration changes have been calculated (up to four open nodes).**

***Probabilities***

**Probabilities for configuration changes are calculated here. Once again different configuration changes will have different probabilities of occurring. All configuration change probabilities have been calculated.**
**Transition matrix is created.**
**Diagonalization is implemented to calculate final probability of induced spark with beginning state of one open channel.**
**Results are presented and plotted.**

***Canonical Form***

**Canonical Form is used to estimate the number of steps it would take to reach our final configuration. Time for each step (delta_time) can be multiplied with this number to estimate time it takes to reach this final configuration.**
**For more information on canonical forms and absorbing Markov chains in general, see [https://en.wikipedia.org/wiki/Absorbing_Markov_chain (https://en.wikipedia.org/wiki/Absorbing_Markov_chain)](https://en.wikipedia.org/wiki/Absorbing_Markov_chain)**

In [1]:
```
import pandas as pd
import numpy as np
import scipy.linalg as la
import scipy as sp
from numpy import trapz
import math
from math import e
import matplotlib.pyplot as plt
```

## Global Variables

In [2]:
```
NEAREST_NEIGHBOR_DISTANCE = 30 #distance of nearest neighbor RyRs from
one another (nanometers)
N = 81 #total numbers of nodes
DIMX = 15
DIMY = 26
DISTANCES_NM = np.linspace(0,130,14)
CALCIUM_CONCENTRATIONS = [25,50,75,100,125,150,175,200,225,250,275,30
0,325,350,375,400,450,500,550,600,650,700,750,800,850,900,950,1000]
STARTING_SINGLE_CHANNEL = sp.array([0,1,0,0,0])
```

```
<ipython-input-2-d69713050b08>:7: DeprecationWarning: scipy.array is
deprecated and will be removed in SciPy 2.0.0, use numpy.array instea
d
  STARTING_SINGLE_CHANNEL = sp.array([0,1,0,0,0])
```

In [3]:
```
Grid_DimX = 9
Grid_DimY = 9
CENTER = np.ceil(Grid_DimY / 2)
N = Grid_DimX * Grid_DimY
gamma = 0.1138 #units are inverse micro molar
lowercase_lambda = 0.2482 #units are inverse seconds
channel_closing_rate = 117 #units are inverse seconds
delta_time = 7.0e-10 #units are in milliseconds
```

In [4]:
```
calcium_profiles = pd.read_excel('Psi Values.xlsx')
calcium_profiles
```

Out[4]:

| | distance, nm | distance, voxels | 25 | 50 | 75 | 100 | 125 | 150 | |
|---|---|---|---|---|---|---|---|---|---|
| 0 | 0.000000 | 0.000000 | 5.325931 | 10.371009 | 15.912928 | 21.176525 | 26.374906 | 31.685792 | 3 |
| 1 | 10.000000 | 1.000000 | 3.325624 | 6.342755 | 9.904539 | 13.161114 | 16.343168 | 19.653902 | 2 |
| 2 | 14.142136 | 1.414214 | 2.782147 | 5.935447 | 8.271289 | 10.981693 | 14.008860 | 16.383152 | 1 |
| 3 | 20.000000 | 2.000000 | 2.428112 | 4.538772 | 7.206327 | 9.560749 | 11.838954 | 14.250627 | 1 |
| 4 | 28.284271 | 2.828427 | 1.952152 | 4.271868 | 5.773700 | 7.649257 | 9.841142 | 11.382286 | 1 |
| 5 | 30.000000 | 3.000000 | 1.907791 | 3.497582 | 5.639754 | 7.470874 | 9.222466 | 11.114340 | 1 |
| 6 | 40.000000 | 4.000000 | 1.552310 | 2.791583 | 4.566778 | 6.039113 | 7.433465 | 8.967646 | 1 |
| 7 | 42.426407 | 4.242641 | 1.469579 | 3.285906 | 4.317070 | 5.705960 | 7.397651 | 8.468049 | |
| 8 | 50.000000 | 5.000000 | 1.285496 | 2.267135 | 3.759076 | 4.962136 | 6.087507 | 7.353381 | |
| 9 | 56.568542 | 5.656854 | 1.137866 | 2.589059 | 3.311491 | 4.365015 | 5.698555 | 6.459276 | |
| 10 | 60.000000 | 6.000000 | 1.073517 | 1.856024 | 3.114835 | 4.103139 | 5.017986 | 6.067707 | |
| 11 | 70.000000 | 7.000000 | 0.898650 | 1.522245 | 2.581155 | 3.392432 | 4.134029 | 5.004553 | |
| 12 | 70.710678 | 7.071068 | 0.891086 | 2.052404 | 2.559412 | 3.362919 | 4.416736 | 4.960106 | |
| 13 | 80.000000 | 8.000000 | 0.750170 | 1.244015 | 2.125723 | 2.786195 | 3.383857 | 4.099190 | |
| 14 | 84.852814 | 8.485281 | 0.699251 | 1.618696 | 1.971256 | 2.580110 | 3.404768 | 3.790699 | |
| 15 | 90.000000 | 9.000000 | 0.621077 | 1.006887 | 1.727777 | 2.257284 | 2.730935 | 3.309837 | |
| 16 | 98.994949 | 9.899495 | 0.546381 | 1.258790 | 1.499565 | 1.953149 | 2.585338 | 2.855470 | |
| 17 | 100.000000 | 10.000000 | 0.506328 | 0.800452 | 1.372160 | 1.784991 | 2.151173 | 2.606085 | |
| 18 | 113.137085 | 11.313708 | 0.423163 | 0.956953 | 1.116929 | 1.445295 | 1.914470 | 2.098947 | |
| 19 | 110.000000 | 11.000000 | 0.402163 | 0.616776 | 1.047792 | 1.354837 | 1.624759 | 1.965541 | |
| 20 | 120.000000 | 12.000000 | 0.305531 | 0.449415 | 0.745577 | 0.954371 | 1.136982 | 1.369895 | |
| 21 | 127.279221 | 12.727922 | 0.323898 | 0.704687 | 0.806800 | 1.034275 | 1.366183 | 1.487417 | |
| 22 | 130.000000 | 13.000000 | 0.213894 | 0.292893 | 0.458064 | 0.573768 | 0.674509 | 0.804038 | |
| 23 | 141.421356 | 14.142136 | 0.245029 | 0.497677 | 0.559047 | 0.706368 | 0.925035 | 1.000048 | |
| 24 | 155.563492 | 15.556349 | 0.184345 | 0.334068 | 0.367536 | 0.453197 | 0.582052 | 0.624071 | |
| 25 | 169.705627 | 16.970563 | 0.140502 | 0.213387 | 0.228672 | 0.269798 | 0.332267 | 0.351881 | |
| 26 | 183.847763 | 18.384776 | 0.112726 | 0.135822 | 0.140472 | 0.153390 | 0.173139 | 0.179189 | |

27 rows C 30 columns

## Hamiltonian Functions

In [5]:
```
#used to estimate psi values that go beyond our given distance data
def found_slope(calcium_concentration):
    numerator = calcium_profiles[calcium_concentration].iloc[-1] - cal
cium_profiles[calcium_concentration].iloc[-2]
    denominator = calcium_profiles['distance, nm'].iloc[-1] - calcium_
profiles['distance, nm'].iloc[-2]
    return numerator / denominator

def extended_psi(calcium_concentration, extended_distance):
    answer = found_slope(calcium_concentration)*(extended_distance - c
alcium_profiles['distance, nm'].iloc[-1]) + calcium_profiles[calcium_c
oncentration].iloc[-1]
    if answer>= 0:
        return answer
    else:
        return 0
```

In [6]:
```
def inverse_temperature(calcium_concentration):
    return (gamma*psi(calcium_concentration, 30))/4
```

In [7]:
```
def round_down(num):
    if num<0:
        return -np.ceil(abs(num))
    else:
        return np.int(num)
```

In [8]:
```
#calcium_concentration is in micromolars and distance_difference is in nanometers
def psi(calcium_concentration, distance_difference):
    normalized_distance = distance_difference/10
    for i in range(len(calcium_profiles['distance, nm']) - 1):
        if distance_difference >= calcium_profiles['distance, nm'].iloc[i] and distance_difference <= calcium_profiles['distance, nm'].iloc[i+1]:
            interval = calcium_profiles['distance, nm'].iloc[i+1] - calcium_profiles['distance, nm'].iloc[i]
            if np.isclose(distance_difference - calcium_profiles['distance, nm'].iloc[i], 0) or np.isclose(distance_difference - calcium_profiles['distance, nm'].iloc[i+1], 0):
                interval_sliver = 0
            else:
                interval_sliver = distance_difference - calcium_profiles['distance, nm'].iloc[i]
            ratio = interval_sliver / interval
            Y_interval = calcium_profiles.loc[calcium_profiles['distance, nm'] == calcium_profiles['distance, nm'].iloc[i+1], calcium_concentration].iloc[0] - \
                                calcium_profiles.loc[calcium_profiles['distance, nm'] == calcium_profiles['distance, nm'].iloc[i], calcium_concentration].iloc[0]
            answer = calcium_profiles.loc[calcium_profiles['distance, nm'] == calcium_profiles['distance, nm'].iloc[i], calcium_concentration].iloc[0] + Y_interval*ratio
    return answer
```

In [9]:
```
def phi(calcium_concentration, distance_difference): #Distance_difference is in nanometers
    return (psi(calcium_concentration, distance_difference)) / (psi(calcium_concentration, 30))
```

In [10]:
```
def summation(calcium_concentration):
    counter = 0
    for j in range(1,Grid_DimY+1):
        for i in range(1,Grid_DimY+1):
            counter += phi(calcium_concentration, 30*math.hypot(CENTER - i, CENTER - j))
    counter -= phi(calcium_concentration, 30*math.hypot(0, 0))
    return counter
```

In [11]:
```
def magnetic_field(calcium_concentration):
    return ((1/(2*inverse_temperature(calcium_concentration)))*(math.log(lowercase_lambda / channel_closing_rate))) \
            + (summation(calcium_concentration))
```

$$Interaction\ Sum\ = \sum_{y \epsilon \Lambda_b} \phi(|x-y|)\sigma(y)$$

In [12]:
```
def interaction_sum_0to1(calcium_concentration):
    return (-1)*summation(calcium_concentration)
```

In [13]:
```
def interaction_sum_1to2(calcium_concentration):
    return (-1)*summation(calcium_concentration) + 2*phi(calcium_conce
ntration, 30)
```

In [14]:
```
def interaction_sum_2to3_straight(calcium_concentration):
    return (-1)*summation(calcium_concentration) + (2)*phi(calcium_con
centration, 30) + (2)*phi(calcium_concentration, 60)


def interaction_sum_2to3_triangle(calcium_concentration):
    return (-1)*summation(calcium_concentration) + (2)*phi(calcium_con
centration, 30) + (2)*phi(calcium_concentration, math.sqrt(30**2 + 30*
*2))
```

In [15]:
```
def interaction_sum_3straight_to_4straight(calcium_concentration):
    return (-1)*summation(calcium_concentration) + (2)*phi(calcium_con
centration, 30) + (2)*phi(calcium_concentration, 60) + (2)*phi(calcium
_concentration, 90)
```

In [16]:
```
def interaction_sum_3straight_to_4L(calcium_concentration):
    return (-1)*summation(calcium_concentration) + (2)*phi(calcium_con
centration, 30) + \
            (2)*phi(calcium_concentration, math.sqrt(30**2 + 30**2)) +
(2)*phi(calcium_concentration, math.sqrt(30**2+60**2))


def interaction_sum_3straight_to_4T(calcium_concentration):
    return (-1)*summation(calcium_concentration) + (2)*phi(calcium_con
centration, 30) + (4)*phi(calcium_concentration, math.sqrt(30**2 + 30*
*2))


def interaction_sum_3triangle_to_4Box(calcium_concentration):
    return (-1)*summation(calcium_concentration) + (4)*phi(calcium_con
centration, 30) + (2)*phi(calcium_concentration, math.sqrt(30**2 + 30*
*2))


def interaction_sum_3triangle_to_4L(calcium_concentration):
    return (-1)*summation(calcium_concentration) + (2)*phi(calcium_con
centration, 30) +\
            (2)*phi(calcium_concentration, 60) + (2)*phi(calcium_conce
ntration, math.sqrt(30**2+60**2))


def interaction_sum_3triangle_to_4S(calcium_concentration):
    return (-1)*summation(calcium_concentration) + (2)*phi(calcium_con
centration, 30) +\
            (2)*phi(calcium_concentration, math.sqrt(30**2 + 30**2)) +
(2)*phi(calcium_concentration, math.sqrt(30**2+60**2))



def interaction_sum_3triangle_to_4T(calcium_concentration):
    return (-1)*summation(calcium_concentration) + (2)*phi(calcium_con
centration, 30) +\
    (2)*phi(calcium_concentration, math.sqrt(30**2 + 30**2)) +\
    (2)*phi(calcium_concentration, 60)
```

## Probabilities

$$P(\sigma, \sigma^x) = \begin{cases} \Delta t C e^{2\beta(\sum_{y \epsilon \Lambda_b} \phi(|x-y|)\sigma(y)+h)}, & \text{for } \sigma(x) = -1 \\ \Delta t C, & \text{for } \sigma(x) = 1 \end{cases}$$

In [17]:
```
def P_1to0(calcium_concentration):
    return delta_time*channel_closing_rate
```

In [18]:
```
def P_1to2(calcium_concentration):
    return 4*delta_time*channel_closing_rate*\
            e**((2*inverse_temperature(calcium_concentration))*(intera
ction_sum_1to2(calcium_concentration)+magnetic_field(calcium_concentra
tion)))
```

In [19]:
```
def P_1to1(calcium_concentration):
    return 1 - P_1to0(calcium_concentration) - P_1to2(calcium_concentr
ation)
```

In [20]:
```
def P_2to1(calcium_concentration):
    return 2*delta_time*channel_closing_rate
```

In [21]:
```
def P_2to3_straight(calcium_concentration):
    return 2*delta_time*channel_closing_rate*\
            e**((2*inverse_temperature(calcium_concentration))*(intera
ction_sum_2to3_straight(calcium_concentration)+magnetic_field(calcium_
concentration)))

def P_2to3_triangle(calcium_concentration):
    return 4*delta_time*channel_closing_rate*\
            e**((2*inverse_temperature(calcium_concentration))*(intera
ction_sum_2to3_triangle(calcium_concentration)+magnetic_field(calcium_
concentration)))

def P_2to3(calcium_concentration):
    return (1/3)*P_2to3_straight(calcium_concentration) + (2/3)*P_2to3
_triangle(calcium_concentration)
```

In [22]:
```
def P_2to2(calcium_concentration):
    return 1 - P_2to1(calcium_concentration) - P_2to3(calcium_concentr
ation)
```

In [23]:
```
def P_3straight_to2(calcium_concentration):
    return 2*delta_time*channel_closing_rate

def P_3triangle_to2(calcium_concentration):
    return 2*delta_time*channel_closing_rate

def P_3to2(calcium_concentration):
    return (1/3)*P_3straight_to2(calcium_concentration) + (2/3)*P_3tri
angle_to2(calcium_concentration)
```

In [24]:
```
def P_3straight_to_4straight(calcium_concentration):
    return 2*delta_time*channel_closing_rate*\
            e**((2*inverse_temperature(calcium_concentration))*(intera
ction_sum_3straight_to_4straight(calcium_concentration)+magnetic_field
(calcium_concentration)))

def P_3straight_to_4L(calcium_concentration):
    return 4*delta_time*channel_closing_rate*\
            e**((2*inverse_temperature(calcium_concentration))*(intera
ction_sum_3straight_to_4L(calcium_concentration)+magnetic_field(calciu
m_concentration)))

def P_3straight_to_4T(calcium_concentration):
    return 2*delta_time*channel_closing_rate*\
            e**((2*inverse_temperature(calcium_concentration))*(intera
ction_sum_3straight_to_4T(calcium_concentration)+magnetic_field(calciu
m_concentration)))

def P_3triangle_to_4Box(calcium_concentration):
    return delta_time*channel_closing_rate*\
            e**((2*inverse_temperature(calcium_concentration))*(intera
ction_sum_3triangle_to_4Box(calcium_concentration)+magnetic_field(calc
ium_concentration)))

def P_3triangle_to_4L(calcium_concentration):
    return 2*delta_time*channel_closing_rate*\
            e**((2*inverse_temperature(calcium_concentration))*(intera
ction_sum_3triangle_to_4L(calcium_concentration)+magnetic_field(calciu
m_concentration)))

def P_3triangle_to_4S(calcium_concentration):
    return 2*delta_time*channel_closing_rate*\
            e**((2*inverse_temperature(calcium_concentration))*(intera
ction_sum_3triangle_to_4S(calcium_concentration)+magnetic_field(calciu
m_concentration)))

def P_3triangle_to_4T(calcium_concentration):
    return 2*delta_time*channel_closing_rate*\
            e**((2*inverse_temperature(calcium_concentration))*(intera
ction_sum_3triangle_to_4T(calcium_concentration)+magnetic_field(calciu
m_concentration)))

def P_3straight_to_4(calcium_concentration):
    return (2/8)*P_3straight_to_4straight(calcium_concentration) + (4/
8)*P_3straight_to_4L(calcium_concentration) + (2/8)*P_3straight_to_4T
(calcium_concentration)

def P_3triangle_to_4(calcium_concentration):
    return (1/7)*P_3triangle_to_4Box(calcium_concentration) + (2/7)*P_
3triangle_to_4L(calcium_concentration) + (2/7)*P_3triangle_to_4S(calci
um_concentration)\
            + (2/7)*P_3triangle_to_4T(calcium_concentration)

def P_3to4(calcium_concentration):
    return (8/15)*P_3straight_to_4(calcium_concentration) + (7/15)*P_3
```

```
triangle_to_4(calcium_concentration)
```

In [25]:
```
def P_3to3(calcium_concentration):
    return 1 - P_3to2(calcium_concentration) - P_3to4(calcium_concentr
ation)
```

In [26]:
```
def transition_matrix(calcium_concentration):
    return sp.array([[1, 0, 0, 0, 0],\
                     [P_1to0(calcium_concentration), P_1to1(calcium_con
centration), P_1to2(calcium_concentration), 0, 0],\
                     [0, P_2to1(calcium_concentration),P_2to2(calcium_c
oncentration), P_2to3(calcium_concentration), 0],\
                     [0, 0, P_3to2(calcium_concentration),P_3to3(calciu
m_concentration), P_3to4(calcium_concentration)],\
                     [0, 0, 0, 0, 1]])
```

## Diagonalization

In [27]:
```
def P(calcium_concentration):
    return la.eig(transition_matrix(calcium_concentration))[1]
```

In [28]:
```
#returns eignevalues for chosen SR level
def eigenvalues(calcium_concentration):
    return la.eigvals(transition_matrix(calcium_concentration))
```

In [29]:
```
def D(calcium_concentration):
    return sp.array([[la.eigvals(transition_matrix(calcium_concentrati
on))[0],0,0,0,0],\
                      [0,la.eigvals(transition_matrix(calcium_concentra
tion))[1],0,0,0],\
                     [0,0,la.eigvals(transition_matrix(calcium_concentr
ation))[2],0,0],\
                     [0,0,0,la.eigvals(transition_matrix(calcium_concen
tration))[3],0],\
                     [0,0,0,0,la.eigvals(transition_matrix(calcium_conc
entration))[4]]])
```

In [30]:
```
def P_inverse(calcium_concentration):
    return la.inv(P(calcium_concentration))
```

In [31]:
```
def D_raised_to_k(calcium_concentration):
    return sp.array([[0,0,0,0,0],\
                      [0,0,0,0,0],\
                      [0,0,0,0,0],\
                      [0,0,0,1,0],\
                      [0,0,0,0,1]])
```

In [32]:
```
def diagonalization_raised_to_k(calcium_concentration):
    PDk = sp.matmul(P(calcium_concentration),D_raised_to_k(calcium_con
centration))
    return sp.matmul(PDk,P_inverse(calcium_concentration))
```

In [33]:
```
def induced_spark(calcium_concentration):
    return sp.matmul(STARTING_SINGLE_CHANNEL,diagonalization_raised_to
_k(calcium_concentration))
```

In [34]:
```
A_LIST = []
for i in CALCIUM_CONCENTRATIONS:
    print(induced_spark(i)[4])
    A_LIST.append(0.85*induced_spark(i)[4])
```

```
<ipython-input-26-47ed076ce5f5>:2: DeprecationWarning: scipy.array is
deprecated and will be removed in SciPy 2.0.0, use numpy.array instea
d
  return sp.array([[1, 0, 0, 0, 0],\
<ipython-input-32-bf24790d66ae>:2: DeprecationWarning: scipy.matmul i
s deprecated and will be removed in SciPy 2.0.0, use numpy.matmul ins
tead
  PDk = sp.matmul(P(calcium_concentration),D_raised_to_k(calcium_conc
entration))
<ipython-input-32-bf24790d66ae>:3: DeprecationWarning: scipy.matmul i
s deprecated and will be removed in SciPy 2.0.0, use numpy.matmul ins
tead
  return sp.matmul(PDk,P_inverse(calcium_concentration))
```

```
2.3082661588788414e-07
6.297227683058175e-07
1.8896870651922238e-06
5.281658616451008e-06
1.5120866008486096e-05
4.0465585738139625e-05
0.000121520155609652 9
0.0003637413025686484
0.0012825347168188338
0.002500824422603198
0.006756177147785917 5
0.014409766759265422
0.02808358687757737
0.035591580710025446
0.06957902157607902
0.1357833718598471
0.2359144874010327
0.3576908093760188
0.48554735049003017
0.6072943213554631
0.7183760693154762
0.7923821416922303
0.852729215006272 4
0.9082186301128848
0.9341159526061776
0.9587479482882124
0.9736392699896748
0.9830954859791495
```

```
In [35]: plt.plot(CALCIUM_CONCENTRATIONS, A_LIST, 'r')
         plt.show()
```

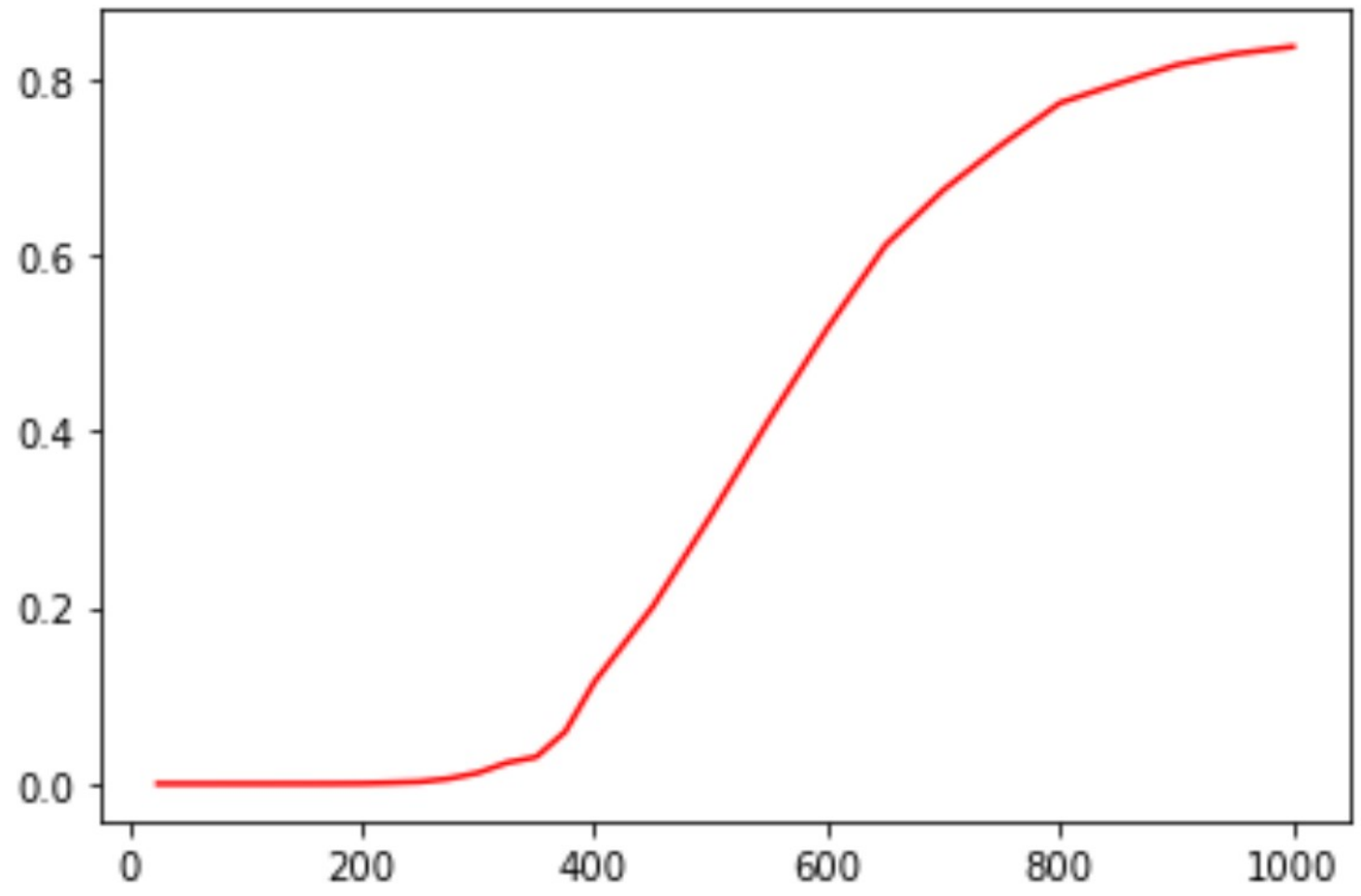

## Canonical Form

```
In [36]: def Q_matrix(calcium_concentration):
             return sp.array([[P_1to1(calcium_concentration), P_1to2(calcium_co
         ncentration), 0],\
                             [P_2to1(calcium_concentration),P_2to2(calcium_conc
         entration), P_2to3(calcium_concentration)],\
                             [0, P_3to2(calcium_concentration), P_3to3(calcium_
         concentration)]])
```

```
In [37]: def R_matrix(calcium_concentration):
             return sp.array([[P_1to0(calcium_concentration), 0],\
                             [0, 0],\
                             [0, P_3to4(calcium_concentration)]])
```

```
In [38]: def Zero_matrix(calcium_concentration):
             return np.zeros([2,3])
```

```
In [39]: def Identity_matrix_r(calcium_concentration):
             return sp.array([[1, 0],[0,1]])
```

```
In [40]: def Identity_matrix_t(calcium_concentration):
             return sp.array([[1,0,0],[0,1,0],[0,0,1]])
```

```
In [41]: def fundamental_matrix(calcium_concentration):
             place_holder_matrix = sp.subtract(Identity_matrix_t(calcium_concen
         tration), Q_matrix(calcium_concentration))
             return la.inv(place_holder_matrix)
```

In [42]:
```
def B_matrix(calcium_concentration):
    return sp.matmul(fundamental_matrix(calcium_concentration), R_matr
ix(calcium_concentration))
```

In [43]:
```
def rate_0to1(calcium_concentration):
    return channel_closing_rate*\
            e**((2*inverse_temperature(calcium_concentration))*(intera
ction_sum_0to1(calcium_concentration)+magnetic_field(calcium_concentra
tion)))
```

In [44]:
```
def rate_1to0(calcium_concentration):
    return channel_closing_rate
```

In [45]:
```
def test_function(calcium_concentration):
    return 71*rate_0to1(calcium_concentration) - rate_1to0(calcium_con
centration)
```

In [46]:
```
def rate_1to2(calcium_concentration):
    return 4*channel_closing_rate*\
            e**((2*inverse_temperature(calcium_concentration))*(intera
ction_sum_1to2(calcium_concentration)+magnetic_field(calcium_concentra
tion)))
```

In [47]:
```
def test_function_2(calcium_concentration):
    return 71*rate_0to1(calcium_concentration) - rate_1to2(calcium_con
centration)
```

In [48]:
```
def test_function_3(calcium_concentration):
    return 71*rate_0to1(calcium_concentration) - (rate_1to2(calcium_co
ncentration) + rate_1to0(calcium_concentration))
```

In [49]:
```
ONE_MATRIX = sp.array([[1],[1],[1]])
```

```
<ipython-input-49-ed93bf915216>:1: DeprecationWarning: scipy.array is
deprecated and will be removed in SciPy 2.0.0, use numpy.array instea
d
  ONE_MATRIX = sp.array([[1],[1],[1]])
```

In [50]:
```
def expected_number_of_steps(calcium_concentration):
    return sp.matmul(fundamental_matrix(calcium_concentration),ONE_MAT
RIX)
```

In [ ]:
```
jupyter nbconvert --to FORMAT notebook.ipynb
```

| distance, nm | distance, voxels | 25 | 50 | 75 | 100 | 125 | 150 | 175 | 200 | 225 | 250 | 275 | 300 | 325 | 350 | 375 | 400 | 450 | 500 | 550 | 600 | 650 | 700 | 750 | 800 | 850 | 900 | 950 | 1000 |
|---|---|---|---|---|---|---|---|---|---|---|---|---|---|---|---|---|---|---|---|---|---|---|---|---|---|---|---|---|---|
| 0 | 0 | 5.325931 | 10.37101 | 15.91293 | 21.17653 | 26.37491 | 31.68579 | 37.09278 | 42.49858 | 47.88034 | 53.09035 | 58.18968 | 63.82965 | 69.07729 | 73.88859 | 79.10741 | 85.44501 | 95.9429 | 106.6517 | 117.4956 | 128.3644 | 139.4183 | 149.7892 | 160.3217 | 171.7891 | 181.8825 | 193.0821 | 203.9759 | 214.8342 |
| 10 | 1 | 3.325624 | 6.342755 | 9.904539 | 13.16111 | 16.34317 | 19.6539 | 23.06247 | 26.47654 | 29.85948 | 33.05663 | 36.13652 | 39.79866 | 43.03742 | 45.79395 | 49.00535 | 53.43114 | 59.90449 | 66.6078 | 73.45833 | 80.3441 | 87.42804 | 93.76002 | 100.2694 | 107.8204 | 113.8506 | 121.0903 | 127.9925 | 134.8506 |
| 14.14213562 | 1.414213562 | 2.7821473 | 5.935447 | 8.271289 | 10.98169 | 14.00886 | 16.38315 | 19.24845 | 22.1183 | 26.10377 | 27.60698 | 31.45144 | 33.2596 | 35.94971 | 36.50042 | 40.81227 | 44.71478 | 50.09278 | 55.68235 | 61.52908 | 67.25131 | 71.81537 | 77.36062 | 81.46592 | 87.65969 | 95.30649 | 100.5011 | 107.3363 | 113.042 |
| 20 | 2 | 2.428112 | 4.538772 | 7.206327 | 9.560749 | 11.83895 | 14.25063 | 16.75737 | 19.27215 | 21.75711 | 24.04677 | 26.22881 | 28.98546 | 31.31814 | 33.15169 | 35.46508 | 39.01177 | 43.67196 | 48.57678 | 53.59487 | 58.67379 | 63.98289 | 68.51001 | 73.20537 | 78.94884 | 83.15491 | 88.58459 | 93.69833 | 98.75811 |
| 28.28427125 | 2.828427125 | 1.9521519 | 4.271868 | 5.7737 | 7.649257 | 9.841142 | 11.38229 | 13.40436 | 15.44048 | 18.59796 | 19.25336 | 22.26597 | 23.22904 | 25.0807 | 24.76274 | 28.26948 | 31.32643 | 35.01973 | 38.93073 | 43.12537 | 47.10639 | 50.02054 | 53.88401 | 56.38941 | 60.82379 | 66.76779 | 70.75611 | 75.4241 | 79.44084 |
| 30 | 3 | 1.907791 | 3.497582 | 5.639754 | 7.470874 | 9.222466 | 11.11434 | 13.08864 | 15.08065 | 17.03877 | 18.8026 | 20.47069 | 22.68703 | 24.49485 | 25.80024 | 27.59818 | 30.60093 | 34.20263 | 38.04989 | 41.99305 | 46.01488 | 50.27358 | 53.74584 | 57.37865 | 62.06624 | 65.21783 | 69.55734 | 73.62645 | 77.61496 |
| 40 | 4 | 1.55231 | 2.791583 | 4.566778 | 6.039113 | 7.433465 | 8.967646 | 10.57384 | 12.20268 | 13.79978 | 15.2018 | 16.52595 | 18.35739 | 19.80365 | 20.75684 | 22.20293 | 24.8064 | 27.68242 | 30.80684 | 33.98615 | 37.26815 | 40.80826 | 43.57115 | 46.47527 | 50.39069 | 52.83031 | 56.40879 | 59.75208 | 63.00826 |
| 42.42640687 | 4.242640687 | 1.4695789 | 3.285906 | 4.31707 | 5.70596 | 7.397651 | 8.468049 | 9.987927 | 11.53265 | 14.15747 | 14.36323 | 16.84543 | 17.34904 | 18.71133 | 17.96137 | 20.94744 | 23.45673 | 26.1633 | 29.08604 | 32.30949 | 35.23054 | 37.22828 | 40.10086 | 41.76243 | 45.10083 | 49.9439 | 52.93537 | 56.57163 | 59.58399 |
| 50 | 5 | 1.285496 | 2.267135 | 3.759076 | 4.962136 | 6.087507 | 7.353381 | 8.676188 | 10.03044 | 11.35115 | 12.48334 | 13.55388 | 15.08517 | 16.26158 | 16.96353 | 18.14808 | 20.41764 | 22.74693 | 25.31714 | 27.92041 | 30.64025 | 33.61728 | 35.84353 | 38.19591 | 41.5232 | 43.4377 | 46.43004 | 49.20472 | 51.8891 |
| 56.56854249 | 5.656854249 | 1.1378658 | 2.58906 | 3.311491 | 4.365015 | 5.698555 | 6.459277 | 7.624811 | 8.825102 | 11.02948 | 10.97514 | 13.04424 | 13.26842 | 14.294 | 13.33352 | 15.89862 | 17.97763 | 20.0041 | 22.23862 | 24.77891 | 26.94589 | 28.35226 | 30.53495 | 31.68541 | 34.22282 | 38.207 | 40.49712 | 43.38369 | 45.68963 |
| 60 | 6 | 1.073517 | 1.856024 | 3.114835 | 4.103139 | 5.017986 | 6.067707 | 7.162757 | 8.293254 | 9.392983 | 10.31041 | 11.18402 | 12.46591 | 13.42663 | 13.94132 | 14.91696 | 16.89487 | 18.79105 | 20.92082 | 23.0501 | 25.31019 | 27.83353 | 29.64443 | 31.56036 | 34.37846 | 35.88848 | 38.41529 | 40.71595 | 42.94723 |
| 70 | 7 | 0.89865 | 1.522245 | 2.581155 | 3.392432 | 4.134029 | 5.004553 | 5.907253 | 6.8505 | 7.763294 | 8.506273 | 9.219882 | 10.28857 | 11.0732 | 11.44784 | 12.25202 | 13.9592 | 15.49948 | 17.2574 | 18.99858 | 20.87442 | 23.00084 | 24.46793 | 26.02342 | 28.4118 | 29.60012 | 31.72326 | 33.62342 | 35.46547 |
| 70.71067812 | 7.071067812 | 0.89108613 | 2.052404 | 2.559412 | 3.362919 | 4.416736 | 4.960106 | 5.855554 | 6.793102 | 8.631941 | 8.433529 | 10.14877 | 10.20202 | 10.97774 | 9.950369 | 12.13569 | 13.84603 | 15.36839 | 17.08447 | 19.09797 | 20.69555 | 21.69457 | 23.35876 | 24.17972 | 26.09422 | 29.35038 | 31.10918 | 33.40112 | 35.17063 |
| 80 | 8 | 0.75017 | 1.244015 | 2.125723 | 2.786195 | 3.383857 | 4.09919 | 4.837289 | 5.616464 | 6.368688 | 6.964755 | 7.544825 | 8.425491 | 9.060415 | 9.328823 | 9.985872 | 11.4399 | 12.6813 | 14.12293 | 15.52778 | 17.06806 | 18.84762 | 20.03184 | 21.28554 | 23.27664 | 24.20557 | 25.97287 | 27.52881 | 29.04158 |
| 84.85281374 | 8.485281374 | 0.69925059 | 1.618696 | 1.971256 | 2.58011 | 3.404768 | 3.790699 | 4.472005 | 5.199128 | 6.70569 | 6.44162 | 7.839629 | 7.794729 | 8.377428 | 7.383794 | 9.209051 | 10.59124 | 11.72583 | 13.03421 | 14.61854 | 15.77708 | 16.48566 | 17.7442 | 18.34293 | 19.76215 | 22.37813 | 23.71758 | 25.51653 | 26.86217 |
| 90 | 9 | 0.621077 | 1.006887 | 1.727777 | 2.257284 | 2.730935 | 3.309837 | 3.902424 | 4.535927 | 5.144675 | 5.616141 | 6.080932 | 6.793625 | 7.300154 | 7.489327 | 8.018433 | 9.227781 | 10.21255 | 11.3732 | 12.49126 | 13.73613 | 15.19429 | 16.13313 | 17.12671 | 18.76238 | 19.478 | 20.91782 | 22.17172 | 23.38827 |
| 98.99494937 | 9.899494937 | 0.54638052 | 1.25879 | 1.499565 | 1.953149 | 2.585338 | 2.85547 | 3.363744 | 3.917537 | 5.117809 | 4.842192 | 5.95122 | 5.858688 | 6.289005 | 5.398785 | 6.878169 | 7.964981 | 8.795745 | 9.775821 | 10.99905 | 11.81866 | 12.31588 | 13.25045 | 13.6918 | 14.71666 | 16.76322 | 17.76454 | 19.14601 | 20.14998 |
| 100 | 10 | 0.506328 | 0.800452 | 1.37216 | 1.784991 | 2.151173 | 2.606085 | 3.06898 | 3.568707 | 4.047976 | 4.410684 | 4.773703 | 5.333133 | 5.725963 | 5.855654 | 6.269595 | 7.242796 | 8.003493 | 8.913471 | 9.775941 | 10.75213 | 11.91386 | 12.64118 | 13.40853 | 14.70513 | 15.24365 | 16.38244 | 17.36518 | 18.31931 |
| 113.137085 | 11.3137085 | 0.42316273 | 0.956953 | 1.116929 | 1.445295 | 1.91447 | 2.098947 | 2.466658 | 2.87584 | 3.795406 | 3.544244 | 4.390946 | 4.285346 | 4.594256 | 3.851384 | 5.001822 | 5.82438 | 6.415565 | 7.128517 | 8.043556 | 8.604189 | 8.945706 | 9.619576 | 9.943048 | 10.65737 | 12.19982 | 12.92631 | 13.95239 | 14.67912 |
| 110 | 11 | 0.402163 | 0.616776 | 1.047792 | 1.354837 | 1.624759 | 1.965541 | 2.30972 | 2.685252 | 3.044084 | 3.31091 | 3.581373 | 3.999398 | 4.290443 | 4.376319 | 4.685305 | 5.426354 | 5.987239 | 6.665849 | 7.302417 | 8.032265 | 8.909575 | 9.446127 | 10.01133 | 10.99153 | 11.37961 | 12.23484 | 12.96829 | 13.67863 |
| 120 | 12 | 0.305531 | 0.449415 | 0.745577 | 0.954371 | 1.136982 | 1.369895 | 1.604073 | 1.861366 | 2.106928 | 2.286678 | 2.47111 | 2.756182 | 2.953383 | 3.006386 | 3.216894 | 3.730069 | 4.109051 | 4.572182 | 5.0013 | 5.499225 | 6.103917 | 6.46757 | 6.849448 | 7.52206 | 7.779749 | 8.366111 | 8.866333 | 9.351103 |
| 127.2792206 | 12.72792206 | 0.32389796 | 0.704687 | 0.8068 | 1.034275 | 1.366183 | 1.487417 | 1.741612 | 2.030293 | 2.697813 | 2.492547 | 3.105455 | 3.008955 | 3.221298 | 2.647091 | 3.492721 | 4.083328 | 4.486277 | 4.982246 | 5.634769 | 6.001127 | 6.227338 | 6.692161 | 6.922544 | 7.397155 | 8.500942 | 9.004851 | 9.730886 | 10.23372 |
| 130 | 13 | 0.213894 | 0.292893 | 0.458064 | 0.573768 | 0.674509 | 0.804038 | 0.933677 | 1.077072 | 1.213575 | 1.312575 | 1.414969 | 1.573156 | 1.682236 | 1.710116 | 1.826778 | 2.113957 | 2.323016 | 2.579884 | 2.816744 | 3.093512 | 3.430671 | 3.631355 | 3.841885 | 4.217564 | 4.357997 | 4.684257 | 4.9618 | 5.230295 |
| 141.4213562 | 14.14213562 | 0.24502927 | 0.497677 | 0.559047 | 0.706368 | 0.925035 | 1.000049 | 1.164201 | 1.354117 | 1.802933 | 1.652979 | 2.064172 | 1.988999 | 2.12559 | 1.721429 | 2.295734 | 2.689103 | 2.946306 | 3.268648 | 3.701754 | 3.926055 | 4.067234 | 4.367025 | 4.521571 | 4.816229 | 5.549532 | 5.876083 | 6.354345 | 6.679413 |
| 155.5634919 | 15.55634919 | 0.18434474 | 0.334068 | 0.367536 | 0.453197 | 0.582052 | 0.624071 | 0.719228 | 0.831084 | 1.099669 | 1.004588 | 1.250256 | 1.200682 | 1.279662 | 1.029865 | 1.376125 | 1.609704 | 1.757455 | 1.945419 | 2.202269 | 2.326351 | 2.406052 | 2.579741 | 2.672888 | 2.837609 | 3.272271 | 3.462192 | 3.743941 | 3.932582 |
| 169.7056275 | 16.97056275 | 0.14050225 | 0.213387 | 0.228672 | 0.269798 | 0.332267 | 0.351881 | 0.39744 | 0.451677 | 0.583307 | 0.534867 | 0.655136 | 0.629262 | 0.667022 | 0.542038 | 0.71253 | 0.826314 | 0.896601 | 0.987041 | 1.112171 | 1.169461 | 1.206855 | 1.290015 | 1.336207 | 1.412682 | 1.624211 | 1.715401 | 1.851951 | 1.942352 |
| 183.8477631 | 18.38477631 | 0.11272644 | 0.135822 | 0.140472 | 0.15339 | 0.173139 | 0.179189 | 0.193468 | 0.210631 | 0.25251 | 0.236735 | 0.274985 | 0.26644 | 0.278271 | 0.238166 | 0.292447 | 0.328482 | 0.350437 | 0.378894 | 0.41857 | 0.43616 | 0.447679 | 0.473764 | 0.48871 | 0.51209 | 0.579086 | 0.607739 | 0.650895 | 0.679247 |

Psi Values.xlsx